# Revisiting the secondary eclipses of KELT-1b using TESS observations

Mohammad Eftekhar[a], Yousefali Abedini[a,b]

[a]*Department of Physics, Faculty of Science, Universityof Zanjan, Zanjan, P. O. Box 313 - 45195, Iran*
[b]*Center for Research on Climate Change and Global Warming, IASBS, Zanjan, Iran*

## ARTICLE INFO

## ABSTRACT

We present the characterization of the transiting planet KELT-1b using data from the Transiting Exoplanet Survey Satellite (TESS). Our light curve model includes primary transit and secondary eclipse. Here, we model the systematic noise using Gaussian processes (GPs) and fit it to the data using the Markov Chain Monte Carlo (MCMC) method. Modelling **of** the TESS light curve returns a planet-to-star radius ratio, $p = 0.07652^{+0.00029}_{-0.00028}$ and a relatively large secondary eclipse depth of $388^{+12}_{-13}$ ppm. The transit ephemeris of KELT-1b is updated using the MCMC method. Finally, we complement our work by searching for transit timing variations (TTVs) for KELT-1b. We do not find significant variations from the constant-period models in our transit time data.

## 1. Introduction

Exoplanets are one of the most exciting and rapidly advancing topics **in** astronomy today. Specifically, transiting planets are valuable because many exoplanet candidates have been detected today by photometric transits. They provide information that may be used to compute the planet's properties, such as its radius and orbital parameters. Since August 2018, the Transiting Exoplanet Survey Satellite (TESS, Ricker et al., 2015) has been delivering high-precision photometric observations **of** a large sample of bright stars from the southern and northern hemispheres. The recorded light curves have given us a treasure trove **of information about** the exoplanet system (Wong et al., 2020; Eftekhar, 2022).

The exoplanet KELT-1b was discovered with the Kilodegree Extremely Little Telescope (KELT) Pepper et al. (2012). Its bright host F5-type star ($V = 10.7$), short orbital period, and inflated radius ($a/R_s = 3.630$, $R_p = 1.15 R_J$) make it one of the best targets for investigating the secondary eclipse depth and ephemeris of its secondary eclipse. Several studies have measured the KELT-1b's primary transit (when an exoplanet passes in front of its host star) and the depth of its secondary eclipse (i.e., when an exoplanet is occulted by its host star) Siverd et al. (2012); Beatty et al. (2019); Kane et al. (2021). Two separate investigations, von Essen et al. (2020) and Beatty et al. (2020) used TESS data to study the full-orbit phase curve of KELT-1b. The planet-to-star radius ratio and eclipse depth of KELT-1b were measured to $304^{+75}_{-75}$ ppm and $0.07688^{+0.00040}_{-0.00040}$ respectively, in the TESS bandpass (von Essen et al., 2020).

Because of its short orbital period, KELT-1b is thought to be tidally locked to its host star (Guillot et al., 1996). According to theoretical calculations and observations, massive exoplanets in tight orbits must decay due to tidal dissipation within their host stars (Maciejewski et al., 2018). Studying precise transit timing allows us to search **for** this orbital evolution. In our study, we're looking for short-term TTVs in sector 17 that might indicate the presence of a third body in this system.

The paper is organized as follows; in Section 2, we describe the TESS observations, data preparation techniques, and our approach to **accounting** for correlated noise. We express our selecting model for primary transit, secondary eclipse, the regression analysis, and TTV in detail in Section 3. We summarize our results from this work in Section 4.

## 2. Observation

KELT-1 (TIC 432549364) was observed by the TESS mission in Sector 17 , Camera 2 , included in the list of preselected target stars observed at a 2-minute cadence. For the results presented in this paper, we decided to use presearch data conditioning (PDC) light curves because they are cleaner than simple aperture photometry light curves (SAP) and show significantly less reduced scatter and short-timescale flux variations (Smith et al., 2012; Stumpe et al., 2014). We also ran a parallel analysis with the light curves from SAP, which were not corrected for systematics by the SPOC pipeline. Our best-fit parameters from SAP light curve analysis are consistent with the $1\sigma$ level with the results from PDC light curves, and most parameters **lie** within $0.2\sigma - 0.3\sigma$.

Although the dominant systematics in the PDCSAP light curve were corrected by default, we corrected it further for the remaining systematics. To do this, we used the median detrending technique with a window length of one orbital period to smooth the PDCSAP light curve, keeping variability **within** the planetary period. This regression was implemented using the Python package `wotan` as shown in Figure 1 (Hippke et al., 2019).

### 2.1. Correlated noise treatment

In our work, we focus on the GP method for modelling correlated noise. Although there are more general definitions of GP, we find that it is suitable for our purposes to describe it as an ordered collection of random variables along one or more axes that often represent time or space. This study models a series of the star's flux, which is taken at discrete

m.eftekhar@znu.ac.ir (M. Eftekhar)
ORCID(s): 0000-0003-1596-0197 (M. Eftekhar)

**Table 1**
The GP model parameters, priors, and best fitted values. The notation $\mathcal{N}(\mu, \sigma^2)$ corresponds to a normal distribution of mean $\mu$ and variance $\sigma^2$, $\mathcal{U}(a, b)$ corresponds to a uniform distribution of lower bound $a$ and higher bound $b$, and $\mathcal{J}(a, b)$ corresponds to a log-uniform distribution between $a$ and $b$. Unit of values is ppm.

| Parameters | Symbol | Prior | Value |
| --- | --- | --- | --- |
| Amplitude of GP | $\sigma_{GP}$ | $\mathcal{J}(10^{-6}, 10^6)$ | $0.00046^{+0.00003}_{-0.00004}$ |
| mean out-of-transit flux | $m_{flux}$ | $\mathcal{N}([0, 0.1])$ | $-0.00016^{+0.00006}_{-0.00006}$ |
| additive photometric jitter term | $\sigma_\omega$ | $\mathcal{J}(10^{-6}, 10^6)$ | $0.01438^{+0.00083}_{-0.00086}$ |
| Matern time-scale | $\rho_{GP}$ | $\mathcal{J}(10^{-3}, 10^3)$ | $0.13553^{+0.01613}_{-0.01872}$ |

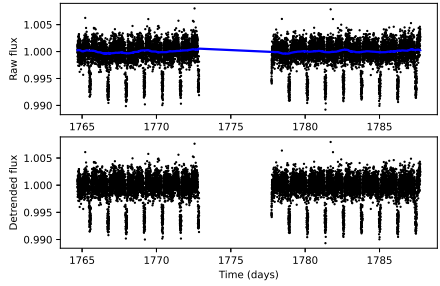

**Figure 1:** (Top) PDCSAP flux of KELT-1 indicated with black dots, and the solid blue line shows the trend obtained by wotan. (Bottom) PDCSAP flux after normalization by its median detrending.

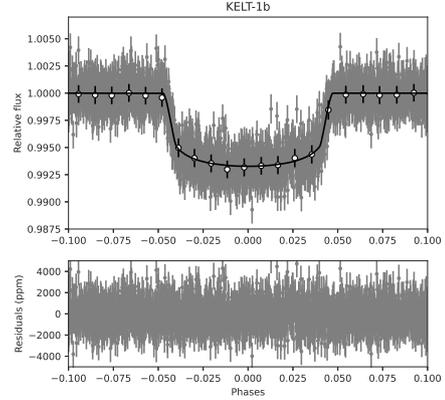

**Figure 2:** Phase-folded light curve presented as grey points showing the primary transit. The binned data (hollow black circle) are overplotted and the best-fitted model (black lines). Corresponding residuals are shown in bottom panel.

times by these random variables. The performance of the GP model for Jupiter- and Neptune-size planets are much better than the non-GP model that gives biased results (Barros et al., 2020).

After masking both the primary transits and the secondary eclipses, we employed a GPs model with a Matern kernel using the celerite package Foreman-Mackey et al. (2017) to detrend the light curve. The uncertainties in the final parameters based on the MCMC approach were calculated using dynamic nested sampling as implemented in dynesty (Speagle, 2020; Skilling, 2004, 2006; Higson et al., 2019). In our further analysis, we will consider this reprocessed data. Table 1 summarizes the prior settings we adopted as well as the best-fit value of each parameter and their uncertainties.

## 3. Analysis
### 3.1. Primary transit modeling

All the planetary parameters in this study were determined using the Juliet (Espinoza et al., 2019). Juliet allows us to fit our photometric data by batman package (Kreidberg, 2015). We determine the posterior probability distribution of the system parameters using the dynamic nested sampling approach, which was implemented in dynesty (Speagle, 2020; Skilling, 2004).

The following parameter transformations were used for the transit model. We set gaussian priors for the orbital period, $P$, and mid-transit time, $T_0$ based on von Essen et al. (2020). Juliet applies the new parametrizations $r_1$, $r_2$ instead of fitting directly for the planet-to-star radius ratio, $p = R_p/R_s$ and the impact parameter of the orbit $b$, for which we set uniform priors between 0 and 1, which ensure uniform sampling of the $b - p$ plane. In addition, instead of individual values of $a/R_s$, we can fit stellar density for all the transiting planets in the system, $\rho_s$, as reported in Tables 2 and 3. We consider a quadratic limb darkening law for our data with a uniform prior between 0 and 1 on both coefficients $q_1$ and $q_2$ (Kipping, 2013). Our model includes a dilution factor, $D_{TESS}$, to represent the ratio of our target's out-of-transit flux to that of all the stars inside the TESS aperture, which has not been corrected for. Due to the fact **that we are** using PDCSAP of TESS (which in principle should have been corrected for the light dilution), we fix the dilution factor to one, which implies no external contaminating sources. The eccentricity, $e$, is also fixed to zero and set to non-informative log-uniform prior to stellar density. We fitted the instrumental jitter term to account for additional systematics and the out-of-transit flux. The combined reprocessed TESS light curve of KELT-1b along with the best-fit model are shown in Figure 2.

The median and $1\sigma$ uncertainties derived from the posterior distributions of our analysis are listed in Table 2. The corner plot for our retrieved posterior distributions from the transit is also shown in Figure 3.

**Table 2**
Prior settings and the best-fit values along with the 68% credibility intervals in the primary transit fit for KELT-1b.

| Parameters | Symbol | Prior | Value |
| --- | --- | --- | --- |
| orbital period(days) | $P$ | $\mathcal{N}(1.21749, 0.1)$ | $1.21749^{+0.00003}_{-0.00004}$ |
| mid-transit time(days) | $T_0$ | $\mathcal{N}(1765.5338, 0.1)$ | $1765.5338^{+0.00032}_{-0.00031}$ |
| parametrisation for $p$ and $b$ | $r_1$ | $\mathcal{U}(0,1)$ | $0.47551^{+0.00880}_{-0.00860}$ |
| parametrisation for $p$ and $b$ | $r_2$ | $\mathcal{U}(0,1)$ | $0.26964^{+0.00029}_{-0.00025}$ |
| quadratic limb darkening coefficient | $q_1$ | $\mathcal{U}(0,1)$ | $0.26964^{+0.00854}_{-0.00859}$ |
| quadratic limb darkening coefficient | $q_2$ | $\mathcal{U}(0,1)$ | $0.29034^{+0.00029}_{-0.00025}$ |
| orbital eccentricity | $e$ | fix | 0 |
| argument of periapsis (deg) | $\omega$ | fix | 90 |
| stellar density ($kg m^{-3}$) | $\rho_s$ | $\mathcal{J}(100, 10000)$ | $604.02^{+7.94}_{-7.84}$ |
| dilution factor | $D_{TESS}$ | fixed | 1 |
| mean out-of-transit (Instrumental offset) | $M_{TESS}$ | $\mathcal{N}(0, 0.1)$ | $0.000004^{+0.000012}_{-0.000013}$ |
| additive photometric jitter term(ppm) | $\sigma_\omega$ | $\mathcal{J}(10^{-6}, 10^{-6})$ | $0.01438^{+0.00091}_{-0.00088}$ |

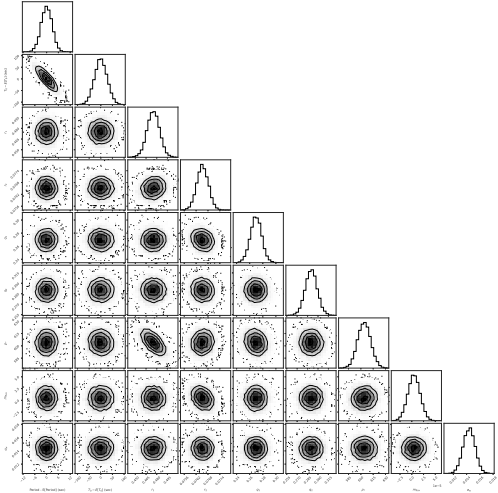

**Figure 3:** Retrieved posterior distributions obtained from our fitting model to the primary transit of the KELT-1b.

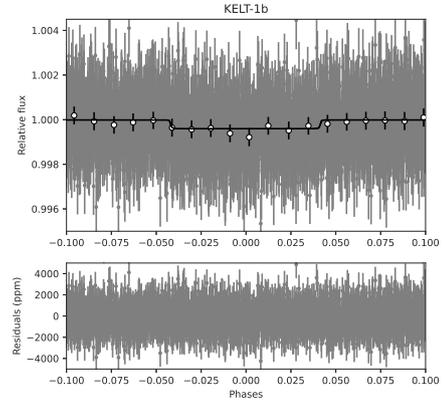

**Figure 4:** A phase-folded light curve is presented as grey points, showing the secondary eclipse. The binned data (hollow black circle) are overplotted and the best-fitted model (black lines). Corresponding residuals are shown in the bottom panel.

### 3.2. Secondary eclipse modeling

Both our transit and eclipse models used the `Juliet`. We use the mid-transit time to calculate the mid-secondary eclipse time for KELT-1b, assuming a circular orbit. The secondary eclipse model is based on the same orbital elements as the primary transit§ 3.1. Therefore, all parameters are coupled to the values of the primary transit except for limb darkening, which fixes $q_1$ to zero, since the secondary eclipse is not affected by limb darkening (Huber et al., 2017). Figure 4 shows our reprocessed data, as well as the best-fitted model of KELT-1b. The results of secondary eclipse model fitting are shown in Table 3.

### 3.3. Transit timing variations

TTV can be used to detect new exoplanets in the system that have a gravitational interaction (Holman and Murray, 2005). In reported results in Table 2, we consider periodic transit events, in other words, the transit times are assumed to be periodic. At this stage, we provide an investigation if our target generates any signature of TTV. For this reason, we fit individual primary transit for each transit time $T_n$ directly. All steps perform, and priors determine as described in the previous section, except $T_0$ and $P$. We set Gaussian priors for each time of transit with a standard deviation of 0.1 days. So, these parameters are directly computed from each sample. This regression is performed using juliet (Espinoza et al., 2019). Figure 5 shows the difference between observed-computed diagrams (O-C) of transit events, which shows a significantly small TTV and no periodic variation in the data.

### 4. SUMMARY AND CONCLUSIONS

In this paper, we characterized the transiting ultra-hot Jupiter KELT-1b. First, we used the median detrending technique with a window length of one orbital period of KELT-1b to conduct smooth detrending on the TESS data. After masking both the primary transits and the secondary eclipses of KELT-1b, we applied the GP model to model correlated noise. With a fitting transit model to reprocessed data, we reliably measure the planetary radius (in stellar radii), ($R_p/R_s$), of $0.07652^{+0.00029}_{-0.00028}$. We measure reasonably large secondary eclipse depth with amplitudes of $388^{+13}_{-12}$ ppm, which is the most precise estimate for KELT-1b to date **Because of our different model compared to the ones used in the previous works**, it's also well within $1\sigma$ of the value of $388^{+67}_{-65}$ ppm reported in the Wong et al. (2021) and significantly ($6.4\sigma$) larger than the measured value of von Essen et al. (2020) of $304 \pm 75$ ppm.

**Table 3**
Prior settings and the best-fit values along with the 68% credibility intervals in the secondary eclipse fit for KELT-1b.

| Parameters | Symbol | Prior | Value |
|---|---|---|---|
| orbital period(days) | $P$ | $\mathcal{N}(1.21749, 0.1)$ | $1.21754^{+0.00021}_{-0.00018}$ |
| mid-eclipse time(days) | $T_{0,e}$ | $\mathcal{N}(1766.74755, 0.1)$ | $1766.1388^{+0.00032}_{-0.00031}$ |
| parametrisation for $p$ and $b$ | $r_1$ | $\mathcal{U}(0, 1)$ | $0.64774^{+0.00090}_{-0.00088}$ |
| parametrisation for $p$ and $b$ | $r_2$ | $\mathcal{U}(0, 1)$ | $0.01976^{+0.00074}_{-0.00073}$ |
| quadratic limb darkening coefficient | $q_2$ | $\mathcal{U}(0, 1)$ | $0.49649^{+0.00920}_{-0.00891}$ |
| orbital eccentricity | $e$ | fix | 0 |
| argument of periapsis (deg) | $\omega$ | fix | 90 |
| stellar density ($kg m^{-3}$) | $\rho_s$ | $\mathcal{J}(100, 10000)$ | $521.73^{+43.94}_{-44.84}$ |
| dilution factor | $D_{TESS}$ | fixed | 1 |
| mean out-of-transit (Instrumental offset) | $M_{TESS}$ | $\mathcal{N}(0, 0.1)$ | $0.00001^{+0.00001}_{-0.00001}$ |
| additive photometric jitter term(ppm) | $\sigma_\omega$ | $\mathcal{J}(10^{-6}, 10^{-6})$ | $0.01433^{+0.00088}_{-0.00089}$ |

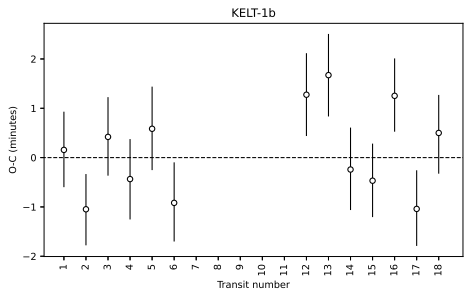

**Figure 5:** TTV amplitudes are calculated in minutes.

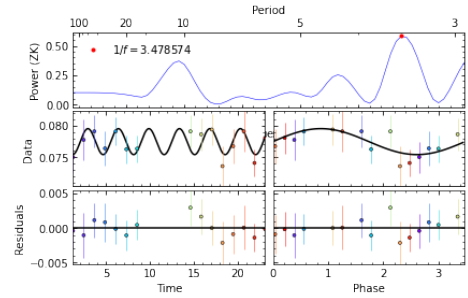

**Figure 6:** GLS periodogram of TTV of KELT-1b shows a clear peak at 3.47 days.

The measured value of the orbital parameters $a/R_s$ and $i$ of $86.6^{+0.2}_{-0.2}$ and $3.646^{+0.0161}_{-0.0158}$, respectively, and they are also the most precise to date and are consistent within $1\sigma$ with the value derived by von Essen et al. (2020) and Beatty et al. (2020). Our estimation **of** the limb darkening coefficients using the Equation, represented by Kipping (2013), is 0.301 and 0.217, for $u_1$ and $u_2$, respectively, which are comparable to those values reported by Claret (2017), equal to limb darkening coefficients of $u_1 = 0.3079$ and $u_2 = 0.2295$. We obtained an updated mid-transit time and the orbital period is consistent with the results of von Essen et al. (2020); Beatty et al. (2020). We find that our results generally lie in good agreement in comparison with other published values in the literature von Essen et al. (2020); Beatty et al. (2020).

The most remarkable result of our study is the most precise detection of the secondary eclipse of KELT-1b in the TESS bandpass and the robust measurement of its orbital parameters. To complement our study, we searched for the individual transit times to look for TTVs. We obtained the O-C diagrams of TTV (see Figure 5), with a standard deviation of 0.87 minutes for KELT-1b, which is significantly small. We also checked if there is any signature of periodicity in measured TTVs using the generalized Lomb-Scargle (GLS) periodogram (Zechmeister and Kürster, 2009). The GLS periodogram on TTV of KELT-1b has been shown in Figure 6, which shows the value of the strongest peak in GLS periodograms is at 3.47 days. For comparison, the stellar rotation rate for our selected host star is $P = 1.33 \pm 0.06$ days based on the values reported in Siverd et al. (2012). We found that our measured TTVs, show periodicity, which was close to the harmonic of the stellar rotation. This can suggest that the variation we measured in TTVs is most probably caused by imperfect elimination of the stellar activity (Oshagh et al., 2013).

## 5. Acknowledgements

NASA's Science Mission Directorate funding the TESS mission. Our work is based on data collected by this mission, available at Mikulski Archive for Space Telescopes (MAST). Special thanks to Mahmoudreza Oshagh, who helped with useful suggestions that improved the paper.